\documentclass[letterpaper]{article}

\usepackage[T1]{fontenc}

\usepackage{geometry}
\usepackage{setspace}

\usepackage[style = chem-acs, autocite = superscript, articletitle=true]{biblatex}
\usepackage{graphicx}
\usepackage{float}
\newfloat{scheme}{htbp}{los}
\floatname{scheme}{Scheme}
\floatname{chart}{Chart}
\newfloat{graph}{htbp}{loh}

\usepackage{chemformula} 
\usepackage[version = 4]{mhchem} 

\usepackage{authblk}

\begin{document}

\title{Mechanisms governing photon-pair generation and emission directionality in quantum metasurfaces}

\author[1,2,3]{Alberto Paniate\thanks{Corresponding author: alberto.paniate@mpl.mpg.de}}
\author[1]{Ivano Ruo-Berchera\thanks{Equal last author contribution.}}
\author[4]{Francesco Monticone\textsuperscript{\(\dagger\)}}

\affil[1]{Quantum Metrology and Nanotechnologies Division, INRiM, Strada delle Cacce 91, 10135 Torino, Italy}
\affil[2]{Friedrich-Alexander-Universität Erlangen-Nürnberg, Staudtstr. 7, 91058 Erlangen, Germany}
\affil[3]{Max Planck Institute for the Science of Light, Staudtstr. 2, 91058 Erlangen, Germany}
\affil[4]{School of Electrical and Computer Engineering, Cornell University, Ithaca, New York 14853, USA}

\date{}

\maketitle

\begin{abstract} 

Metasurfaces are emerging as a promising platform for photon-pair generation through spontaneous parametric down-conversion, thanks to their compactness, integrability, and intrinsic multifunctionality, which enables the engineering of complex quantum states. However, their full potential remains only partially exploited because the physical mechanisms governing key properties of the generated photon pairs, such as generation efficiency and emission directionality, are not yet fully understood. As a result, metasurface designs and experimental configurations are often optimized through trial-and-error procedures. Here, we theoretically investigate the main mechanisms that control photon-pair generation and detection by studying how different pump configurations and measurement geometries affect the generation efficiency, emission directionality, and collection efficiency of the emitted photon pairs. This framework allows us to interpret existing experimental results and to provide general guidelines for the design of metasurfaces and the choice of experimental configurations in future experiments. Finally, we show that substrate thickness and multilayer configurations represent additional degrees of freedom for quantum metasurface design and can be engineered to enhance the generation efficiency and control the emission directionality, providing a new route for the optimization of photon-pair sources based on metasurfaces.
\end{abstract}	


\maketitle
\section{Introduction}
Traditionally, spontaneous parametric down-conversion (SPDC) has been performed in bulk non-centrosymmetric crystals, where the millimeter-scale interaction length between the pump field and the nonlinear medium enables efficient generation of entangled photon pairs, commonly referred to as signal and idler photons. However, this long interaction length also imposes stringent phase-matching constraints, thereby limiting the accessible properties of the generated photon pairs and restricting the choice of suitable dispersive materials. In recent years, micrometer- and nanometer-thick nonlinear films have attracted considerable interest as compact and integrable SPDC sources with relaxed longitudinal phase-matching condition \autocite{okoth2019microscale, santiago2021entangled, lu2025, sultanov2022flat, stich2026thin, liang2025}.  Nevertheless, the intrinsically low conversion efficiency of ultra-thin nonlinear media has motivated the development of structured nonlinear systems, in which the resonances supported provide strong field enhancement and thereby significantly increase the photon-pair generation efficiency. SPDC has been demonstrated both in single nanoresonators \autocite{marino2019spontaneous, duong2022spontaneous, saerens2023background, poloczek2026efficient} and in metasurfaces, i.e. periodic arrays of nanoresonators \autocite{santiago2021photon, zhang2022spatially, santiago2022resonant, ma2023polarization, son2023photon, noh2024quantum, weissflog2024directionally, noh2025fano, ma2025quantum, jia2025polarization, ma2025nonlinearity}. 

Beyond efficiency enhancement, nonlinear metasurfaces offer a large number of design degrees of freedom. This flexibility enables multifunctional quantum-light sources with properties that are difficult to achieve in bulk crystals, such as engineered spatial emission \autocite{zhang2022spatially, son2023photon, weissflog2024directionally}, tunable and arbitrary polarization Bell-state generation \autocite{ma2023polarization, ma2025nonlinearity, jia2025polarization}, and cluster-state generation \autocite{santiago2022resonant}. Although their efficiencies remain generally lower than those of conventional bulk sources, the multifunctional properties achievable, together with compactness and integrability, represent their main advantage for quantum light generation.

However, the full exploitation of this multifunctionality requires accurate design and optimization strategies tailored to specific target properties. At present, this progress is strongly limited by the lack of a comprehensive theoretical framework describing the mechanisms that govern quantum emission from nonlinear metasurfaces. As a result, key photon-pair properties, such as generation efficiency and emission directionality, as well as the optimal collection geometry, are often difficult to predict and are still largely determined through trial-and-error approaches.

Here, we take an important step forward in this direction by theoretically investigating, within a quasi-normal-mode (QNM) framework \autocite{poddubny2016generation, weissflog2024nonlinear, poloczek2026efficient}, how different pump directions and detection geometries affect the photon-pair generation efficiency and the collection of the emitted photons. The theoretical predictions are compared with existing experimental results obtained from a lithium-niobate (LN) metasurface fabricated on a glass substrate \autocite{santiago2021photon}, providing a physical interpretation of the observed behavior.

Motivated by the crucial role of substrate-induced effects emerging from this analysis, we investigate both the generation efficiency and the emission directionality first within a semi-infinite substrate approximation, which isolates the intrinsic response of the metasurface by neglecting finite-thickness effects, and then for substrates with finite thickness and multiple layers. By addressing these effects, which remain largely unexplored in SPDC metasurfaces, our analysis opens new routes to enhance photon-pair generation and tailor the emission properties of quantum metasurfaces, similarly to recent substrate-interference strategies used to enhance second-harmonic generation in two-dimensional materials \autocite{song2023interference, puri2024substrate}.

\section{Results and Discussion}
\subsection{Theoretical framework and mode structure}

\begin{figure}[t]
    \centering
    \includegraphics[width=0.6\linewidth]{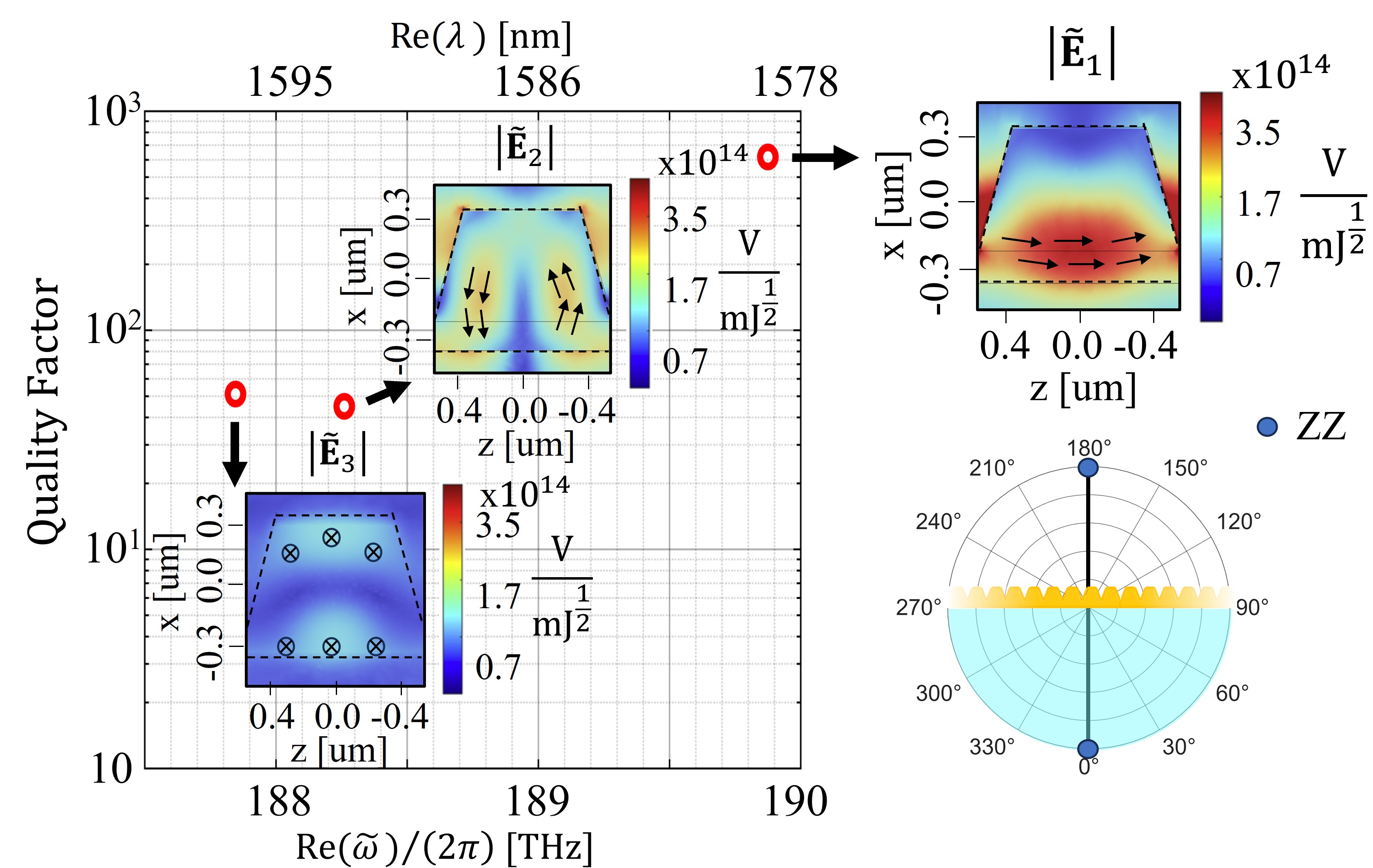}
    \caption{ {\textit{Quasi-normal modes (QNMs) supported by the metasurface.} 
    The metasurface, deposited on a glass substrate, supports three QNMs, characterized by complex eigenfrequencies \(\tilde{\omega}=\omega-i\gamma\), where \(\omega\) is the resonance frequency and \(\zeta\) is the leakage rate. The QNMs are indicated by red dots, together with their resonance frequencies and quality factors, defined as \(Q=\omega/(2\zeta)\). The insets show the corresponding electric near-field distributions inside a single unit-cell of the infinite metasurface, with black arrows and crossed circles indicating the local electric-field polarization. The right panel highlights the dominant mode, labeled \(\mathrm{QNM}_1\), which exhibits the strongest electric field inside the nonlinear material. Its near-field distribution is shown in the upper panel, while the corresponding far-field emission pattern is shown in the lower panel as a function of the polar angle.}}
    \label{fig:transmission}
\end{figure}

To illustrate the proposed theoretical framework we consider a realistic example. The investigated metasurface consists of an infinite periodic array, with periodicity of 900 \(\text{nm}\), of truncated pyramids (the complete geometrical and nonlinear parameters are provided in section~1 of the Supplementary Material) and is excited by a pump field at wavelength \(\lambda_p = 788~\text{nm}\) \autocite{santiago2021photon}. The metasurface supports several resonances, shown in Fig.~\ref{fig:transmission} and labeled with numbers 1-3. Owing to their intrinsically leaky nature, these resonances are commonly referred to as QNMs \autocite{kristensen2014modes,lalanne2018light}. The red dots in Fig.~\ref{fig:transmission} indicate the different QNMs together with their corresponding quality factors, defined as \(Q = \omega /(2\,\zeta)\), where \(\omega\) and \(\zeta\) are respectively the real and imaginary parts of the complex eigenfrequency \(\tilde{\omega} = \omega - i\,\zeta\). Here, \(\omega\) determines the resonance frequency, while \(\zeta\) describes the radiative leakage rate of the mode. 

The QNMs are shown together with their electric near-field distributions inside a single unit-cell of the metasurface \(|\tilde{\mathbf{E}}|\). Each QNM is normalized according to QNMs theory  ~\autocite{sauvan2022normalization} and detailed in Supplementary Material sections 1-2. The black arrows and crossed circles indicate the local polarization of the electric field. Among the different modes, \(\mathrm{QNM}_1\), whose resonance wavelength is closest to the degenerate SPDC wavelength, (\(2\,\lambda_p=1576~\mathrm{nm}\)), exhibits the largest near-field amplitude inside the nonlinear material. Moreover, its electric field is predominantly oriented along the optical axis of LN, corresponding to the \(z\)-direction, for which the nonlinear tensor component \(\chi^{(2)}_{_{zzz}}\) is the largest.

For these reasons, in the following we use QNM$_1$ as a representative resonant channel to construct a reduced description of the SPDC response and isolate the main mechanisms governing the substrate-dependent generation and directionality of photon pairs.

The QNMs shown in Fig.~\ref{fig:transmission} are retrieved with zero Floquet wave-vector, i.e., zero parallel component of the wave-vector of the field, which describes the normal emission channel. The main analysis is restricted to this channel. The validity and limitations of this approximation are examined in Supplementary Material Section 3.

Below the near-field profile of QNM$_1$ in Fig.~\ref{fig:transmission}, the angular dependence of the corresponding complex electric far-field, $|\widetilde{\mathbf{E}}(\theta)|$, is shown.
For the geometry considered here, QNM$_1$ exhibits approximately equal normal far-field electric-field amplitudes in the air and substrate half-spaces. We stress that equal electric-field amplitudes do not imply equal emitted photon fluxes, owing to the different refractive indices of the two surrounding media. The far field is
predominantly $z$-polarized, consistently with the dominant polarization of the QNM near field inside the LN resonators. Therefore, in the following analysis we restrict the discussion to the $z$-polarized contribution of QNM$_1$.

\begin{figure*}
    \centering
    \includegraphics[width=\textwidth]{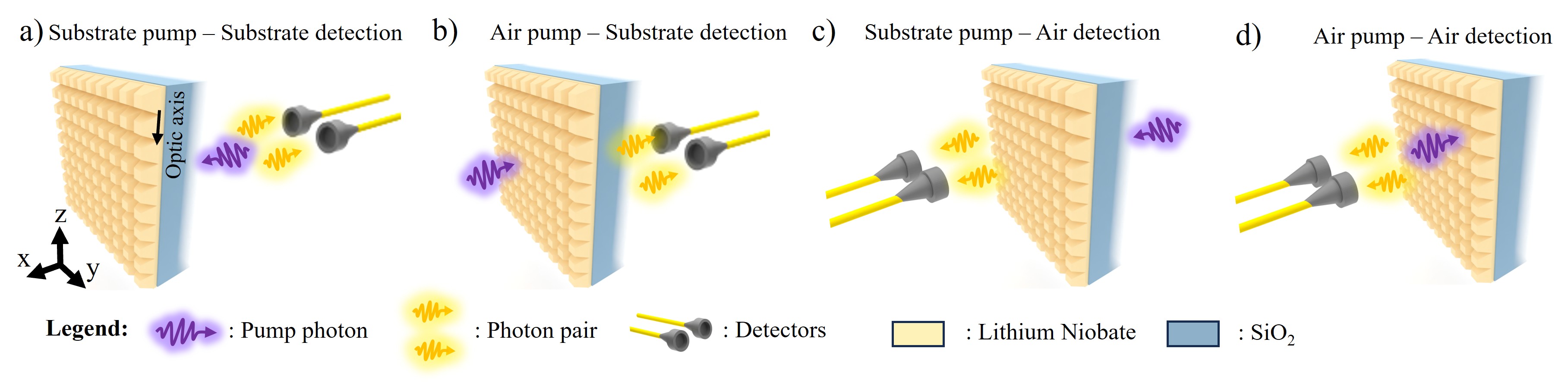}
    \caption{\textit{Schematic representation of the four pump and detection geometries considered in this work.} A lithium-niobate metasurface with the optic axis aligned along the \(z\)-axis is deposited on a glass substrate \autocite{santiago2021photon} and is pumped by a laser under different excitation and collection configurations. (a) Pump incident from the substrate side and photon-pair detection from the substrate side. (b) Pump incident from the air side and detection from the substrate side. (c) Pump incident from the substrate side and detection from the air side. (d) Pump incident from the air side and detection from the air side.}
    \label{fig:configurations}
\end{figure*}

The contribution of the QNM pair \(\mathrm{QNM}_1-\mathrm{QNM}_1\) to the SPDC photon-pair detection rate is given by \autocite{poddubny2016generation,weissflog2024nonlinear, poloczek2026efficient}
\begin{equation}
   |\tilde{T}_{\mathrm{is}}|^2 =  \Bigg|\xi(\omega_\mathrm{s}) \,\tilde{E}_{\mathrm{z}}(\theta_\mathrm{i})\,\tilde{E}_\mathrm{z}(\theta_\mathrm{s})\Bigg|^2 ,
\label{eq:TPA}
\end{equation}
where \(\tilde{T}_{\mathrm{is}}\) is the complex two-photon amplitude, namely the probability amplitude associated with the joint detection of the idler and signal photons. The signal and idler angular frequencies are \(\omega_{\mathrm{s}}\) and \(\omega_{\mathrm{i}}=\omega_{\mathrm{p}}-\omega_{\mathrm{s}}\), respectively, where \(\omega_{\mathrm{p}}\) is the pump angular frequency. The two photons are detected in the far-field along the directions of polar angles \(\theta_\mathrm{i}\) and \(\theta_\mathrm{s}\), respectively, with polarization along the \(z\)-axis. In the normal-emission case considered here, the relevant far-field amplitudes are evaluated along the normal emission channels: \(\theta=0^\circ\) on the substrate side and \(\theta=180^\circ\) on the air side, as shown in Fig.~\ref{fig:transmission}.

While the QNMs are intrinsic properties of the metasurface and therefore independent of the excitation field, the contribution of a QNM pair is weighted by the nonlinear modal overlap coefficient \(\xi(\omega_\mathrm{s})\):

\begin{equation}
    \xi(\omega_\mathrm{s})=\frac{\sum_{\alpha,\beta,\gamma}\int d\mathbf r_0\, \chi^{(2)}_{\alpha\beta\gamma}(\mathbf r_0)\, \tilde E_{\alpha}(\mathbf r_0)\, \tilde E_{\beta}(\mathbf r_0)\, E_{\mathrm{p},\gamma}(\mathbf r_0)}{(\omega_\mathrm{p}-\omega_\mathrm{s}-\tilde\omega)\, \tilde\omega\, (\omega_\mathrm{s}-\tilde\omega)\, \tilde\omega} 
    \label{eq:overlap}
\end{equation}
where the numerator quantifies the nonlinear interaction inside the LN resonators through the spatial overlap between the nonlinear susceptibility tensor \(\chi^{(2)}\), the QNM near-field \(\tilde{\mathbf{E}}\) along directions \(\alpha\) and \(\beta\), and the pump field \(\mathbf{E}_\mathrm{p}\) along direction \(\gamma\). The detuning term in the denominator, instead, accounts for the spectral mismatch between the generated photons \((\omega_\mathrm{s},\,\omega_\mathrm{i})\) and the complex eigenfrequency \(\tilde{\omega}\) of \(\mathrm{QNM}_1\). For the complete model see Supplementary Material section 1.

Within this theoretical framework, we analyze how different excitation and collection geometries affect the detected photon-pair rate.

\subsection{SPDC efficiency and directionality with different excitation-collection geometries and semi-infinite substrate}

We consider a plane-wave excitation incident on the metasurface supported by a semi-infinite glass substrate. The theoretically calculated number of detected photon pairs is integrated over the spectral range from 1550~nm to 1600~nm,  and the resulting count rates are then compared for the four different excitation and detection geometries illustrated in Fig.~\ref{fig:configurations}. 

The metasurface is excited either from the substrate side or from the air side and, for each excitation configuration, photon-pair collection is analyzed both from the air side (\(\theta_\mathrm{s} = \theta_\mathrm{i} = 180^\circ\)) and from the substrate side (\(\theta_\mathrm{s} = \theta_\mathrm{i} = 0^\circ\)). Experimentally, the highest enhancement, corresponding to approximately a factor of \(20\) compared to a thin film of equal thickness, was observed for the reflection configuration (a). By contrast, configurations (b) and (c), referred to as transmission geometries, since the photon pairs are collected along the pump propagation direction, showed no measurable enhancement. Although configuration (d) was not experimentally investigated, our analysis predicts that low enhancement is expected in this case as well, as discussed later.

Importantly, changing the pumping or detection geometry does not modify the intrinsic QNM properties or their eigenfrequencies. Therefore, the difference between the investigated configurations arises from two configuration-dependent factors: the pump-dependent nonlinear overlap \(|\xi|^2\), particularly through the numerator of Eq.~\eqref{eq:overlap}, and the portion of the QNM far-field that is selected by the detection geometry, which enters through the factor \(\left| \tilde{E}_\mathrm{z}(\theta_{\mathrm{s}})
\, \tilde{E}_\mathrm{z}(\theta_{\mathrm{i}})\right|^2.\)

\begin{figure*}
    \centering
    \includegraphics[width=0.8\textwidth]{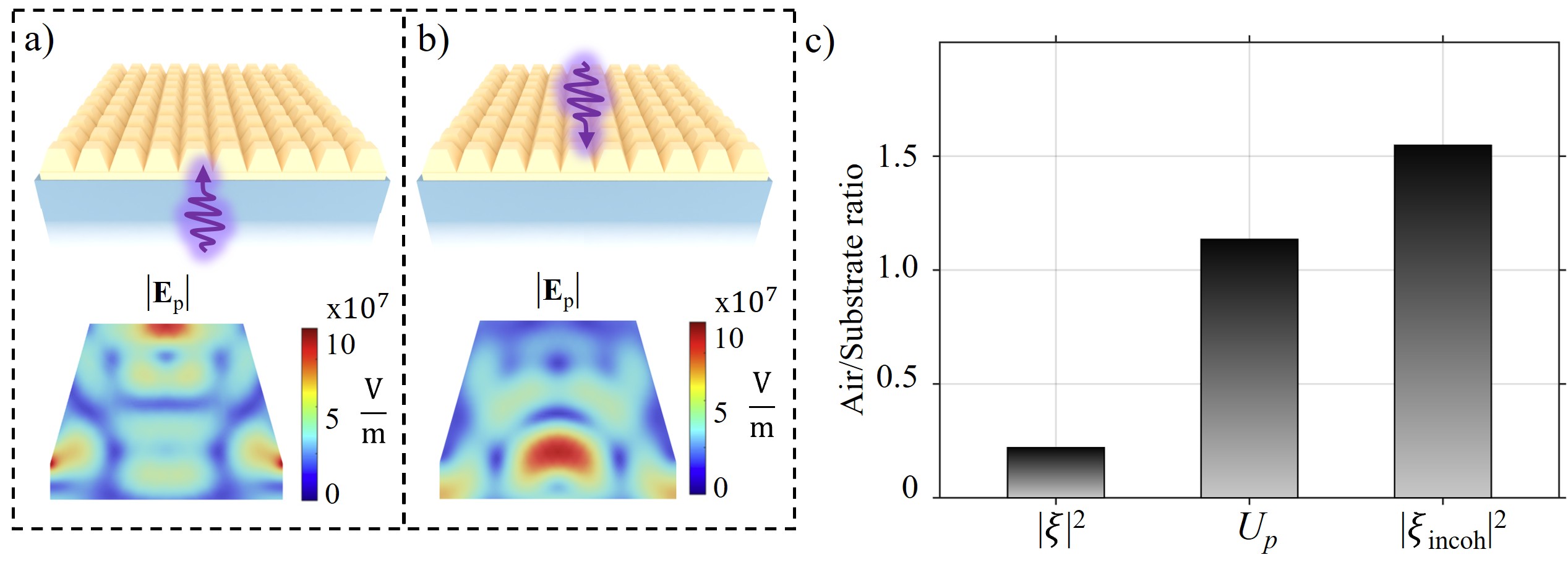}
    \caption{\textit{Effect of the pump direction on the nonlinear modal overlap.}
    (a,b) Pump electric-field amplitude \(|\mathbf{E}_\mathrm{p}|\) inside one unit-cell of the LN metasurface for substrate-side pumping (a) and air-side pumping (b). (c) Air-side/substrate-side ratios of the coherent nonlinear-overlap contribution \(|\xi|^2\), the integrated pump-field intensity inside the nonlinear material \(U_\mathrm{p}\), and the incoherent overlap \(|\xi_{\mathrm{incoh}}|^2\). All values are evaluated for equal incident pump photon flux in the two excitation configurations.}
    \label{fig:substrate}
\end{figure*}

Fig.~\ref{fig:substrate} compares configurations (a) and (b) in which the photon pairs are collected from the same side, while the pump is incident either from the substrate side or from the air side. Since the collection side is unchanged, the QNM far-field contribution in Eq.~\eqref{eq:TPA} remains the same, while the numerator in Eq. (\ref{eq:overlap}) changes through the different pump-field distributions inside the LN resonators. The corresponding pump electric-field amplitudes inside one unit-cell are shown in Fig.~\ref{fig:substrate} (a,b), revealing clear differences between the two excitation configurations. These differences arise from the asymmetry of the resonators and the different pump fields experienced by the metasurface. In fact, for substrate-side excitation, the metasurface is predominantly driven by the field transmitted through the glass substrate. In contrast, for air-side excitation, the pump field inside the LN resonators results from the interference between the incident field and the field reflected and scattered by the metasurface--substrate system. As a consequence, the two pumping configurations generate distinct spatial pump-field distributions inside the nonlinear material.

The effect of the pump direction is summarized in Fig.~\ref{fig:substrate} (c), which reports the air-side/substrate-side ratios of three relevant quantities. The first bar shows the ratio of the modal overlap contribution \(|\xi|^2\).  Air-side pumping yields a value of only \(\sim 0.2\) relative to substrate-side pumping, indicating that substrate-side excitation enhances the SPDC generation efficiency by approximately a factor of five, in qualitative agreement with the experimental observations.

To determine whether this enhancement simply originates from a larger pump intensity inside the nonlinear material, rather than an enhanced modal overlap, the second bar shows the ratio of the integrated pump-field intensity inside the LN resonators, defined as
\begin{equation}
U_{\mathrm{p}} =
\int
|\mathbf E_{\mathrm{p}}(\mathbf r_0)|^2\,d\mathbf r_0 .
\end{equation}
The ratio of \(U_{\mathrm{p}}\) is close to unity, indicating that comparable amounts of pump-field intensity are coupled into the nonlinear material for the two excitation directions.

The third bar reports the ratio of the incoherent overlap, \(|\xi_{\mathrm{incoh}}|^2\), obtained by taking the absolute value of the integrand in Eq.~\eqref{eq:overlap} before performing the spatial integration. This quantity is larger for air-side pumping, indicating that the intensity overlap between the pump field and the nonlinear source distribution associated with \(\mathrm{QNM}_1\) is not reduced in this configuration. This result is also qualitatively consistent with the field distributions shown in Fig.~\ref{fig:substrate} (a,b) and Fig.~\ref{fig:transmission}. In fact, for air-side excitation, the pump-field amplitude exhibits a stronger concentration in the lower central region of the LN resonator, where \(\mathrm{QNM}_1\) also displays a pronounced near-field amplitude.

Therefore, the strong suppression of \(|\xi|^2\) for air-side excitation cannot be attributed either to a lower pump intensity inside the LN resonators or to a weaker intensity overlap. Instead, it originates from stronger phase cancellations within the coherent nonlinear-overlap integral. This demonstrates that the pump direction  controls the phase and symmetry matching between the pump field and the QNM nonlinear source distribution, thereby determining the photon-pair generation efficiency.

We next compare configurations (a) and (c), where the pump is incident from the substrate side in both cases, while the photon pairs are collected either from the substrate side (\(\theta_\mathrm{s} = \theta_\mathrm{i} = 0^\circ\)) or from the air side (\(\theta_\mathrm{s} = \theta_\mathrm{i} = 180^\circ\)). Since the excitation geometry is unchanged, the pump-field distribution inside the LN resonators is the same for the two configurations. Consequently, the  overlap factor \(|\xi|^2\) is also unchanged, and the difference in the detected photon-pair rate originates from the far-field emission and collection factors in Eq.~\eqref{eq:TPA}. 

\begin{figure*}
    \centering
    \includegraphics[width=\textwidth]{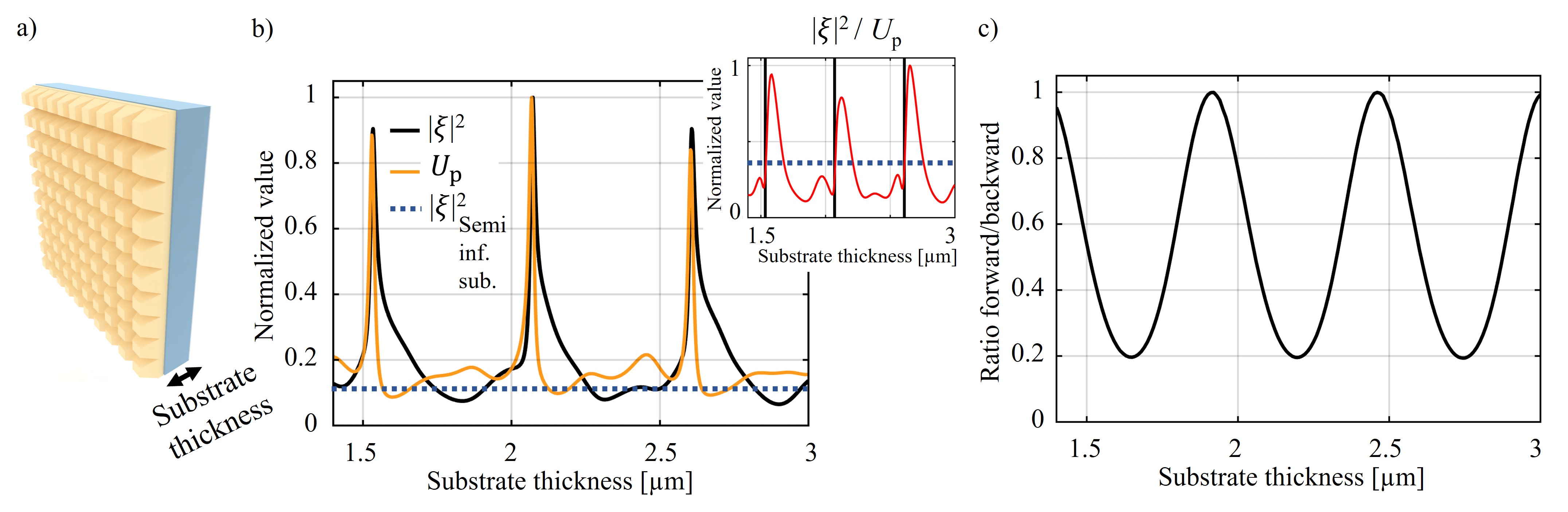}
    \caption{\textit{Finite-substrate effects on SPDC generation and emission directionality.} (a) Schematic of the finite-substrate geometry, where the substrate thickness is varied.  (b) Normalized nonlinear modal overlap contribution, \(|\xi|^2\), and integrated pump-field intensity inside the nonlinear material, \(U_{\mathrm{p}}\), as a function of the substrate thickness. The horizontal blue dashed line indicates the value of \(|\xi|^2\) obtained for a semi-infinite substrate. The inset shows the normalized ratio \(|\xi|^2/U_{\mathrm{p}}\); the black vertical lines mark the substrate thicknesses corresponding to the main \(|\xi|^2\) peaks, while the blue dashed line indicates the semi-infinite-substrate value of the same ratio. All curves are independently normalized to their maximum value. For each layer configuration, the QNMs and their normalizations were calculated independently. (c) Ratio between air and substrate photon-pair emission as a function of the substrate thickness.} 
    \label{fig:finitesubstrate}
\end{figure*}

As shown in Fig.~\ref{fig:transmission}, the far-field electric-field amplitude of \(\mathrm{QNM}_1\), calculated for the metasurface on a semi-infinite glass substrate, is symmetric between the air and substrate half-spaces. However, this polar plot represents the QNM electric-field amplitude, not directly the emitted photon flux. In fact, due to the different refractive indices of the two media, comparable far-field amplitudes do not correspond to equal photon fluxes. The final air-side to substrate-side photon-pair emission ratio is approximately 0.44 (see Supplementary Material section~2). Consequently, configuration~(a) exhibits a higher detected photon-pair rate than configuration~(c), in qualitative agreement with the experimental observations \autocite{santiago2021photon}.

We note, however, that this comparison is based on the infinite-periodic-metasurface approximation. Although Floquet-periodic boundary conditions are commonly used to describe metasurfaces, experiments probe only a finite region of the structure due to the finite pump-beam waist. Finite-aperture effects are known to modify resonant properties such as the quality factor, radiative out-coupling, far-field emission pattern, and edge-related leakage \autocite{droulias2018finite, ustimenko2024resonances}.

In the original experiment ~\autocite{santiago2021photon}, the pump beam had a finite width of approximately \(6~\mu\mathrm{m}\), corresponding to the excitation of only \(7\)--\(8\) unit-cells. To assess the impact of this finite excitation aperture, we simulated a finite array of nanoresonators with a comparable number of coherently excited unit-cells (Supplementary Material section 4). In this case, the far-field emission becomes strongly asymmetric and is predominantly directed toward the substrate, as a result of the combined effect of the finite illuminated area and the refractive-index mismatch between the air and substrate half-spaces, consistent with the known asymmetric emission of finite antennas and quantum emitters above substrates \autocite{engheta2002radiation, curto2010unidirectional}. The resulting air-side to substrate-side photon-pair emission ratio is reduced below \(0.1\), in agreement with the experimental observation of no measurable emission in this configuration.

We emphasize that this finite-array analysis is specific to the experimental conditions considered here, where only a few unit-cells are effectively excited by the focused pump beam. For larger illumination areas, the infinite-periodic approximation is expected to become increasingly accurate. Nevertheless finite-excitation-aperture effects can substantially impact the emission directionality and collection efficiency, and should therefore be carefully considered when drawing quantitative conclusions. Further details are provided in the Supplementary Material section 4.

Finally, although configuration (d) was not investigated experimentally, our model predicts it to yield the lowest efficiency. This is because it combines a reduced nonlinear overlap, \(|\xi|^2\), with inefficient collection due to the weak emission of the mode in the corresponding detection direction.

\subsection{SPDC efficiency and directionality with a finite substrate}

While the previous analysis was performed by considering a semi-infinite substrate, thereby neglecting etalon effects induced by multiple reflections in the substrate, we now consider a finite-thickness substrate, as shown in Fig.~\ref{fig:finitesubstrate} (a), and study its effect on the generation efficiency and emission directionality. In the following, we focus on the case of a plane-wave pump incident from the substrate side, given the higher efficiency found in the previous analysis.

Fig.~\ref{fig:finitesubstrate} (b) shows the nonlinear-overlap contribution \(|\xi|^2\) as a function of the substrate thickness. The response exhibits a sequence of pronounced and narrow maxima, with variations approaching one order of magnitude. In particular, the main efficiency peaks closely follow the behavior of the integrated pump-field intensity inside the nonlinear material, \(U_\mathrm{p}\), shown by the orange curve.

This modulation originates from finite-substrate interference effects. Since the metasurface period is \(900~\mathrm{nm}\), larger than the pump wavelength inside the substrate, the pump field scattered by the metasurface can be reflected into several diffraction orders in the glass. These diffraction orders undergo multiple reflections at the substrate interfaces and interfere constructively or destructively depending on the substrate thickness. As a result, the pump field coupled into the LN resonators becomes strongly thickness-dependent, leading to the observed modulation of \(|\xi|^2\). A complete analytical description of the periodicity and its relation to the relevant diffraction orders is provided in section~5 of the Supplementary Material.

The constructive recoupling of reflected or weakly coupled diffraction orders through multiple interference inside the substrate enables an enhancement by almost one order of magnitude compared to the semi-infinite substrate case, whose efficiency is indicated by the horizontal dashed blue line in Fig.~\ref{fig:finitesubstrate} (b).

Although this enhanced coupling is beneficial for reducing unwanted pump reflection and increasing the nonlinear interaction, the largest efficiency peaks are mainly associated with a larger integrated pump-field intensity inside the LN resonators. This aspect is important for practical operation, since large field intensities confined in the nonlinear material may increase the risk of optical damage, heating, or other intensity-induced degradation mechanisms. It is therefore useful to evaluate not only the absolute overlap contribution \(|\xi|^2\), but also the generation efficiency normalized to the pump-field intensity stored in the nonlinear material. To this aim, the inset in Fig.~\ref{fig:finitesubstrate} (b) shows the ratio \(|\xi|^2 / U_\mathrm{p}\). This quantity isolates the part of the efficiency enhancement that does not simply arise from an increased pump intensity inside the nonlinear material. Indeed, the substrate thickness modifies not only the total amount of pump field coupled into the metasurface, but also its spatial distribution and phase inside the LN resonators, modifying the final generation efficiency.

This mechanism is reflected in the oscillatory behavior of \(|\xi|^2/U_\mathrm{p}\), which exhibits maxima and minima differing by almost one order of magnitude. Moreover, the maxima of this normalized efficiency are slightly shifted with respect to the maxima of \(U_\mathrm{p}\), shown with the black vertical lines. In particular, there is a maximum enhancement of approximately a factor of three compared to the semi-infinite glass substrate case, indicated with blue line. This demonstrates that substrate-thickness engineering can improve the photon-pair generation efficiency by optimizing the coherent nonlinear overlap, rather than simply increasing the pump intensity inside the metasurface.

These results suggest a useful optimization strategy for damage-limited operation: instead of maximizing only \(|\xi|^2\), one can search for substrate thicknesses that maximize the photon-pair generation efficiency per pump-field intensity stored in the nonlinear material. In this way, the SPDC efficiency can be enhanced through improved coherent overlap, without necessarily increasing the local pump-field intensity inside the metasurface.

Finally, Fig.~\ref{fig:finitesubstrate} (c) shows the ratio between photon-pair emission in the air and substrate directions as a function of the substrate thickness. This ratio exhibits pronounced smooth oscillations, originating from Fabry--Pérot-like interference effects inside the finite substrate. Indeed, at the signal and idler wavelengths the metasurface period is smaller than the wavelength in both surrounding media, so that only the zeroth diffraction order is propagating and all higher diffraction orders are evanescent. Furthermore, the modulation is enhanced by the high-\(Q\) resonance of the metasurface, which generates spectrally narrow photon pairs with a correspondingly long coherence length. As a result, the different reflected contributions inside the substrate can interfere coherently and modify the relative emission into the two directions. 

Although the present analysis is limited to a finite glass substrate of a few micrometers surrounded by air, it suggests a more general strategy based on realistic multilayer substrates. To support this perspective, we investigate an elementary multilayer configuration in section~6 of the Supplementary Material, showing that substrate engineering can provide an experimentally feasible route to enhance and tailor SPDC from nonlinear metasurfaces.

\section{Conclusion}

In this work, we have developed a theoretical framework that identifies the main mechanisms that govern photon-pair generation and collection in a nonlinear metasurface and used it to provide a physical interpretation of existing experimental results. Our analysis extends the use of a QNM-based description of SPDC, previously developed for single nanoresonators \autocite{weissflog2024nonlinear, poloczek2026efficient}, to a periodic metasurface supporting high-\(Q\) resonances. Within this framework, the different excitation and collection geometries are understood through two main contributions: the pump-dependent coherent nonlinear modal overlap and the direction-dependent far-field emission of the selected resonant QNM. Furthermore, the present analysis should be regarded as a first step toward a more complete multimode description of SPDC in periodic metasurfaces, in which the contributions of multiple QNMs, together with their angular dispersion and dependence on the Floquet wave vector, are treated consistently in order to reconstruct the full angle-resolved biphoton state.

Our results also highlight the important role of substrate-related effects in SPDC metasurfaces. While finite-substrate effects have so far received little attention in this context, we show that substrate thickness can act as an additional design parameter for controlling both generation efficiency and emission directionality. In particular, finite-substrate interference and diffraction-order recoupling can enhance the pump coupling to the metasurface and modify the coherent nonlinear overlap. Importantly, the efficiency can be improved not only by increasing the pump-field intensity inside the nonlinear material, but also by optimizing the nonlinear overlap for a comparable stored pump intensity. This provides a promising route toward enhanced operation under practical constraints associated with optical damage, heating, or other intensity-induced degradation mechanisms. The proof-of-principle multilayer example presented in the Supplementary Material further indicates that the same physical mechanism can be translated into more realistic substrate geometries, provided that the metasurface period, pump wavelength, and refractive-index environment are designed together.

Although the quantitative enhancement factors reported here are specific to the investigated LN metasurface, the physical mechanisms identified by our analysis suggest general guidelines for resonant SPDC metasurfaces. First, the excitation configuration should be optimized by maximizing the coherent nonlinear overlap. Second, the collection geometry should be selected by considering the QNM far-field emission and the photon flux into the relevant surrounding media, while finite-excitation-aperture effects should be accounted for whenever only a limited number of unit-cells are effectively illuminated. Third, when substrate-induced interference is exploited, the substrate thickness and refractive-index environment should be co-designed with the metasurface period and pump wavelength so that the relevant pump diffraction channels are efficiently recoupled to the nonlinear resonators. Finally, for damage-limited operation, a useful optimization target is the generation efficiency per pump-field intensity stored in the nonlinear material, as quantified here by \(|\xi|^2/U_{\mathrm{p}}\).

These design principles provide a foundation for future optimization and inverse-design strategies \autocite{stich2025inverse} aimed at realizing more efficient, directional, and experimentally robust quantum metasurface sources. 

\section{Associated content}

\subsection{Preprint}
Paniate, A.; Ruo-Berchera, I.; Monticone, F. Mechanisms governing photon-pair generation and emission directionality in quantum metasurfaces. 2026, 2606.26303. arXiv. https://arxiv.org/abs/2606.26303 (accessed September 16, 2026).

\subsection{Supporting information}
Geometrical and nonlinear parameters of the metasurface (S1.1); Theoretical model and QNMs normalization (S1.2); Transmittance retrieved with QNMs (S2); QNMs with non-zero Floquet boundary conditions (S3); Finite size metasurface simulation (S4); Substrate effects induced by diffraction channels reflection (S5); Proof of principle of a multilayers design (S6). 

\section{ACKNOWLEDGMENTS}

A.P. and F.M. thank ISSNAF for enabling the collaboration.
A.P. thanks Prof. Maria Chekhova and Dr. Tomàs Santiago-Cruz for useful discussions.

\subsection{Funding Sources}

A.P. acknowledges funding from the European Research Council (Project 101199215 — MultiFlaQS).
F.M. acknowledges funding from the U.S. National Science Foundation (Grant No. DMR-2522004).

\section{Conflict of interest}
The authors declare no conflicts of interest.

\printbibliography
\end{document}


\begin{center}

{\large\bfseries Supporting Information}

\vspace{0.5cm}

{\large
Mechanisms governing photon-pair generation and emission
directionality in quantum metasurfaces}

\vspace{0.4cm}

Alberto Paniate,$^{1,2,3,*}$
Ivano Ruo-Berchera,$^{1,\dagger}$
and Francesco Monticone$^{4,\dagger}$

\vspace{0.3cm}

$^1$ Quantum metrology and nano technologies division,
INRiM, Strada delle Cacce 91, 10135 Torino, Italy

$^2$ Friedrich-Alexander-Universität Erlangen-Nürnberg,
Staudtstr. 7, 91058 Erlangen, Germany

$^3$ Max Planck Institute for the Science of Light,
Staudtstr. 2, 91058 Erlangen, Germany

$^4$ School of Electrical and Computer Engineering,
Cornell University, Ithaca, New York 14853, USA

\vspace{0.3cm}

$^*$ Corresponding author: alberto.paniate@mpl.mpg.de  $^\dagger$ Equal last author contribution.

\end{center}

\section{Materials and Computational Methods}

\subsection{Geometrical and nonlinear parameters of the metasurface }

\begin{figure}[h]
    \centering
    \includegraphics[width=0.25\linewidth]{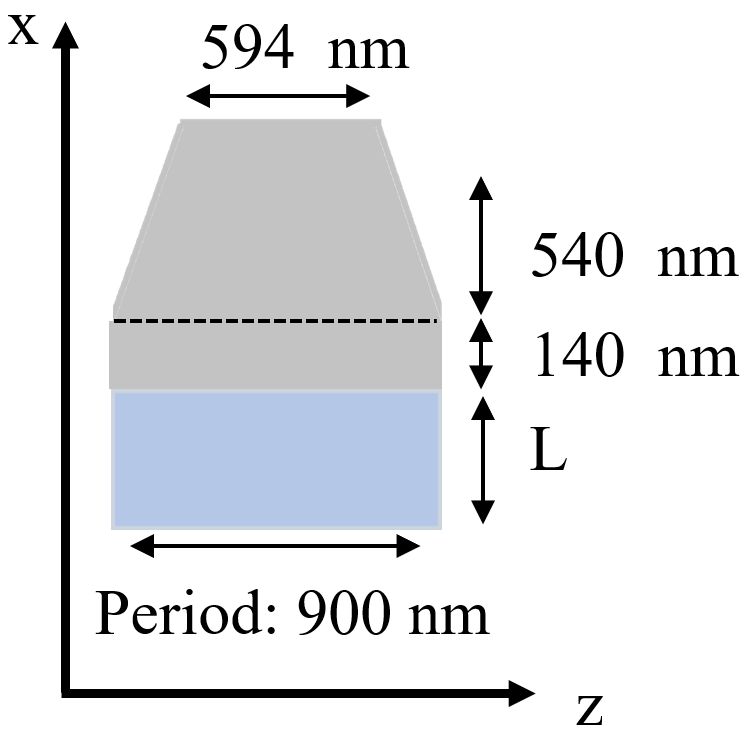}
    \caption{\textit{Geometrical parameters of a single unit-cell of the lithium niobate (LN) metasurface}}
    \label{fig:pyramid}
\end{figure}
In Fig. \ref{fig:pyramid} the unit cell is depicted. It consists of a truncated pyramid with a base length equal to the metasurface period (900 nm), a top base length of 594 nm, and a height of 540 nm, called metasurface B in the original paper \autocite{santiago2021photon}. The truncated pyramids rest on a residual layer with an estimated height of 140 nm, which is supported by a glass substrate of height \(L\). 
The nanostructures are made of lithium niobate (LN), a material well-known for its high second-order nonlinear susceptibility \(\chi^{(2)}\). The polarization response, in the laboratory reference frame, can be described as \autocite{santiago2021photon}:
\begin{equation}
\begin{aligned}
P_x^{\mathrm{NL}}(\omega_p) &= 4\varepsilon_0 \Big[ d_{31}\big(E_x(\omega_s)E_z(\omega_i) + E_z(\omega_s)E_x(\omega_i)\big) - d_{22}\big(E_x(\omega_s)E_y(\omega_i) + E_y(\omega_s)E_x(\omega_i)\big) \Big], \\[0.5em] P_y^{\mathrm{NL}}(\omega_p) &= 4\varepsilon_0 \Big[
d_{22}\big(E_y(\omega_s)E_y(\omega_i) - E_x(\omega_s)E_x(\omega_i)\big) + d_{31}\big(E_y(\omega_s)E_z(\omega_i) + E_z(\omega_s)E_y(\omega_i)\big) \Big], \\[0.5em]
P_z^{\mathrm{NL}}(\omega_p) &= 4\varepsilon_0 \Big[ d_{31}\big(E_x(\omega_s)E_x(\omega_i)
+ E_y(\omega_s)E_y(\omega_i)\big) + d_{33}E_z(\omega_s)E_z(\omega_i) \Big].
\end{aligned}
\label{eq:nonlinear_polarization_components}
\end{equation}
where \(E_{x,y,z}(\omega)\) denote the Cartesian components of either the signal electric field, at angular frequency \(\omega_s\), or the idler electric field, at angular frequency \((\omega_\mathrm{i}=\omega_\mathrm{p}-\omega_\mathrm{s}) \), while \(\omega_p\) is the pump angular frequency. The nonlinear coefficients are \(d_{22}=1.9~\mathrm{pm/V}\), \(d_{31}=-3.2~\mathrm{pm/V}\), and \(d_{33}=\frac{1}{2}\chi_{zzz}^{(2)}=-19.5~\mathrm{pm/V}\) at \(1313~\mathrm{nm}\),
where the contracted notation, \(d\), for \(\chi^{(2)}\) was used \autocite{shoji1997absolute}. 
As a first approximation the LN was considered with two sets of refractive indices, one at the wavelength of the resonance \((\approx 1570\) nm : \((n_{\textrm{x}}, n_{\textrm{y}}, n_{\textrm{z}}) = (2.21, 2.21, 2.14)\) and the other at the pump wavelength (788 nm) : \((n_{\textrm{x}}, n_{\textrm{y}}, n_{\textrm{z}}) = (2.26, 2.26, 2.18)\). The refractive index of the glass substrate is set to \(n_g = 1.44\), and the surrounding background is considered to be air.

\subsection{Theoretical model}
Assuming ideal detection, the photon-pair detection rate can be written as \autocite{poddubny2016generation, weissflog2024nonlinear}:
\begin{equation}
\frac{\text{d}^4 N_{\mathrm{pair}}}{\text{d}t\, \text{d}\Omega_\mathrm{i}\, \text{d}\Omega_\mathrm{s}\, \text{d}\omega_\mathrm{s}}
 = \, \mathcal{C}\left(n_\mathrm{i},\,\omega_\mathrm{p} - \omega_\mathrm{s};\,n_\mathrm{s},\,\omega_\mathrm{s}\right) \,  \left| \widetilde{T}_{\mathrm{is}}\!\left(\theta_\mathrm{i}, \varphi_\mathrm{i},\,\omega_\mathrm{p} - \omega_\mathrm{s},\,\mathbf{d}_\mathrm{i};\,\theta_\mathrm{s}, \varphi_\mathrm{s},\,\omega_\mathrm{s},\,\mathbf{d}_\mathrm{s}\right) \right|^2,
\label{eq:pair_rate}
\end{equation}
where
\begin{equation}
    \mathcal{C}\left(n_\mathrm{i},\,\omega_\mathrm{p} - \omega_\mathrm{s};\,n_\mathrm{s},\,\omega_\mathrm{s}\right) \equiv \frac{8}{\pi \mu_0^2}\, n_\mathrm{i} n_\mathrm{s}\,
\frac{(\omega_\mathrm{p} - \omega_\mathrm{s})^3 \omega_\mathrm{s}^3}{{c_\mathrm{0}}^6}
\end{equation}
and 
\begin{equation}
   \widetilde{T}_{\mathrm{is}} =  \sum_{\mathrm{m,n} = 1}^{\infty}
\xi_{\mathrm{m,n}}(\omega_\mathrm{s})\,
\tilde{E}_{\mathrm{m},\mathbf{d}_\mathrm{i}}(\theta_\mathrm{i}, \varphi_\mathrm{i})\,
\tilde{E}_{\mathrm{n},\mathbf{d}_\mathrm{s}}(\theta_\mathrm{s}, \varphi_\mathrm{s}).
\label{eq:TPA}
\end{equation}
is the complex two-photon amplitude, namely the probability amplitude for the joint detection of the idler and signal photons in the far-field directions specified by the spherical angles \((\theta_\mathrm{i},\varphi_\mathrm{i})\) and \((\theta_\mathrm{s},\varphi_\mathrm{s})\), with detection polarizations \(\mathbf d_\mathrm{i}\) and \(\mathbf d_\mathrm{s}\), respectively. The refractive indices of the surrounding medium at the idler and signal frequencies are denoted by \(n_\mathrm{i}\) and \(n_\mathrm{s}\), whereas \(\mu_0\) and \(c_0\) are the vacuum permeability and the speed of light in vacuum.

The detection rate is given per unit far-field solid angle, \(d\Omega_\mathrm{i}d\Omega_\mathrm{s}\), and per unit signal-photon angular frequency, \(d\omega_\mathrm{s}\). The quantity \(\tilde{E}_{m,\mathbf d_\mathrm{i}}(\theta_\mathrm{i},\varphi_\mathrm{i})\) denotes the angular dependence of the complex far-field electric field of the \(m\)-th QNM, projected onto the detection polarization \(\mathbf d_\mathrm{i}\). The summation runs over all QNM pairs (m,n), with the contribution of each pair weighted by the so-called modal-overlap coefficient
\begin{equation}
\xi_{\mathrm{m,n}}(\omega_\mathrm{s}) \equiv \frac{\mathcal{G}_{m,n}}{\mathcal{S}_{m,n}(\omega_\mathrm{s})},
\end{equation}
where
\begin{equation}
\mathcal{G}_{\mathrm{m,n}} =
\sum_{\mathrm{\alpha,\beta,\gamma}}
\int d\mathbf r_0\,
\chi^{(2)}_{\alpha\beta\gamma}(\mathbf r_0)\,
\tilde E_{\mathrm{m,\alpha}}(\mathbf r_0)\,
\tilde E_{\mathrm{n,\beta}}(\mathbf r_0)\,
E_{\mathrm{p,\gamma}}(\mathbf r_0)
\label{eq:G_factor}
\end{equation}
and
\begin{equation}
\mathcal{S}_{\mathrm{m,n}}(\omega_\mathrm{s}) =(\omega_\mathrm{p}-\omega_\mathrm{s}-\tilde\omega_\mathrm{m})\tilde\omega_\mathrm{m}
(\omega_\mathrm{s}-\tilde\omega_\mathrm{n})\tilde\omega_\mathrm{n}.
\label{eq:overlFactor}
\end{equation}
The integral \(\mathcal{G}_{\mathrm{m,n}}\) quantifies the spatial overlap between the nonlinear susceptibility tensor \(\chi^{(2)}\), the near-field QNM eigenfields \(\tilde{\mathbf{E}}_\mathrm{m}\) and \(\tilde{\mathbf{E}}_\mathrm{n}\) along the directions \(\alpha\) and \(\beta\), and the pump field \(\mathbf{E}_\mathrm{p}\) along the direction \(\gamma\). The spectral factor \(\mathcal{S}_{\mathrm{m,n}}(\omega_\mathrm{s})\) accounts for the complex detuning between the generated photon frequencies \((\omega_\mathrm{i},\omega_\mathrm{s})\) and the QNM eigenfrequencies \((\tilde{\omega}_\mathrm{m},\tilde{\omega}_\mathrm{n})\), thereby determining the spectral and modal selectivity of SPDC. 

The detected photon-pair count rate is calculated  by integrating over all the range of angles allowed by the numerical aperture (NA) of the system and all the frequency 
\begin{equation}
    \frac{\mathrm{d} N_{\mathrm{pair}}}{\mathrm{d}t} =  \int \mathrm{d}\omega_\mathrm{s} \int_{\text{NA}} \mathrm{d}\Omega_\mathrm{i}\, \int_{\text{NA}} \mathrm{d}\Omega_\mathrm{s} \,
    \mathcal{C} \, \left| \widetilde{T}_{\mathrm{is}}\!\left(\theta_\mathrm{i},\, \varphi_\mathrm{i},\,\omega_\mathrm{p} -\omega_\mathrm{s}, \, \mathbf{d}_\mathrm{i}; \,\theta_\mathrm{s},\, \varphi_\mathrm{s}, \,\omega_\mathrm{s}, \,\mathbf{d}_\mathrm{s}\right) \right|^2,
    \label{eq:final}
\end{equation}
where the variables entering the prefactor~$\mathcal{C}$ have been omitted for brevity.

The QNMs are normalized according to the standard PML-based QNM normalization \autocite{lalanne2018light,sauvan2022normalization}. In particular, the electric and magnetic fields of each QNM are rescaled such that
\begin{equation}
\label{eq:QNMnorm}
\langle \tilde{\mathbf{E}}_m | \tilde{\mathbf{E}}_m \rangle =
\int d\mathbf{r}\,
\Big[
\tilde{\mathbf{E}}_m(\mathbf{r}) \cdot 
\frac{\partial (\varepsilon(\omega)\,\omega)}{\partial \omega}\,
\tilde{\mathbf{E}}_m(\mathbf{r}) -\tilde{\mathbf{H}}_m(\mathbf{r}) \cdot
\frac{\partial (\mu(\omega)\,\omega)}{\partial \omega}\,
\tilde{\mathbf{H}}_m(\mathbf{r})
\Big] = 1 ,
\end{equation}
where \(\varepsilon(\omega)\) and \(\mu(\omega)\) represent the permittivity and permeability of the domain, respectively, and the derivatives are evaluated at the complex eigenfrequency \(\tilde{\omega}_m\). \(\tilde{\mathbf{H}}_m(\mathbf{r})\) denote the magnetic-field profiles of the \(m\)-th QNM and the integral is performed over the whole simulated volume including the perfectly matched layers (PML). The correctness of this normalization is proven in the following transmission calculation.\\
In the main text since only the normal emission was considered the far-field \(\widetilde{E}_{z}(\theta=0)\) was calculated as
\begin{equation}
    \widetilde{E}_{z}(\theta=0)
    = \frac{1}{p^2}
    \int_{-p/2}^{p/2} \mathrm{d}y \int_{-p/2}^{p/2} \,\mathrm{d}z\,
    \widetilde{E}_{z}(x_1,y,z),
    \label{eq:NFFF}
\end{equation}
where \(p\) is the size of the unit-cell of the infinite periodical metasurface and $x_1$ denotes the position of the near-to-far transformation plane in the homogeneous air or substrate region.

\section{Transmittance with QNMs}

\begin{figure}[H]
    \centering
    \includegraphics[width=0.9\linewidth]
    {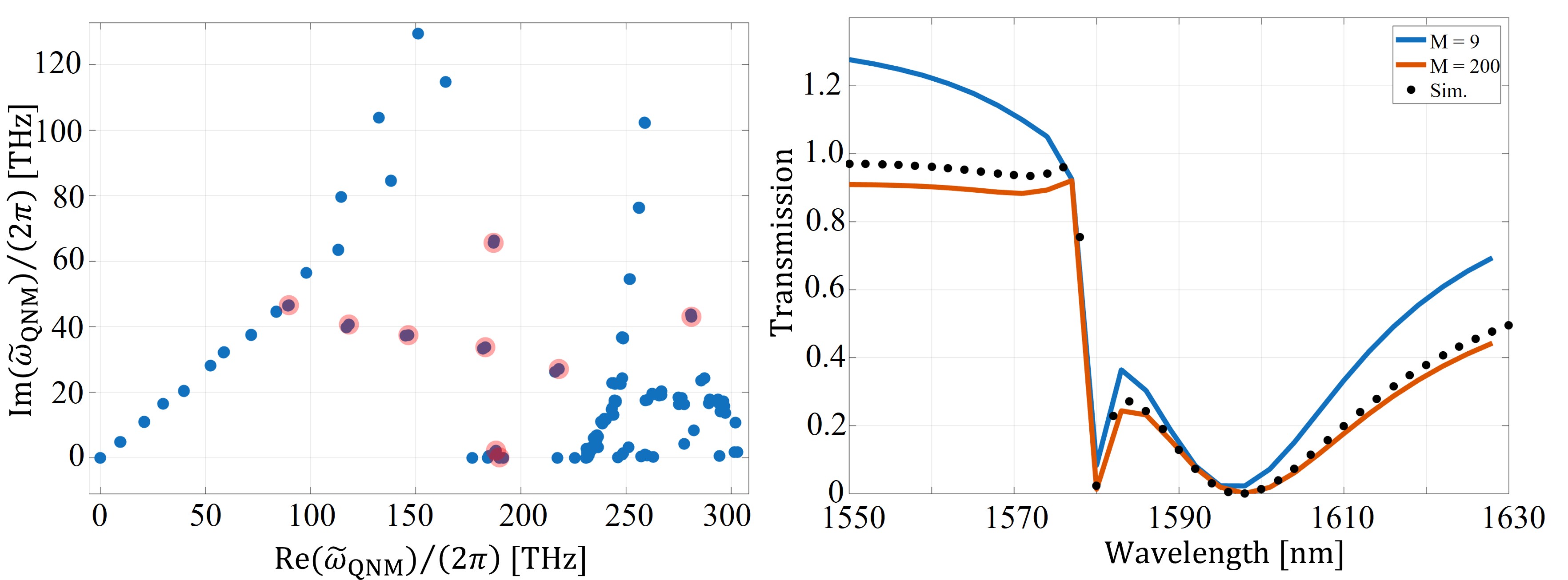}
    \caption{\textit{Validation and convergence of the QNM reconstruction of the linear transmission.}
    Left: complex eigenfrequencies of the QNMs obtained from the eigenfrequency calculation. 
    The nine modes providing the largest contributions to the transmitted field in the considered spectral region are highlighted in red.
    Right: comparison between the transmission obtained from the full-wave simulation (black dots) and from the QNM reconstruction using $M=9$ (blue line) and $M=200$ (orange line) modes.}
    \label{fig:TransmissionQNM}
\end{figure}

Here, we reconstruct the linear transmission spectrum of the periodic metasurface using an increasing number of QNMs, following the QNM expansion formalism \autocite{lalanne2018light}, and compared it with the transmission obtained directly from the full-wave simulation to verify the correctness of QNMs retrieval and normalization. The results are shown in Fig.~\ref{fig:TransmissionQNM}. The left panel shows the complex eigenfrequencies of the QNMs considered in the reconstruction. The subset of nine modes providing the largest contributions to the transmitted field in the spectral range of interest is highlighted in red. The right panel compares the directly simulated transmission (black dots) with QNM reconstructions obtained using $M=9$ modes (blue curve) and $M=200$ modes (orange curve).
As the number of QNMs is increased, the reconstructed transmission progressively approaches the full-field result. In particular, the reconstruction with $M=200$ modes reproduces the position and spectral shape of the main resonant features.

As is generally encountered in QNM expansions of open electromagnetic systems, an accurate reconstruction of the linear response over a broad spectral range may require the inclusion of low-$Q$ numerical modes associated with the PML discretization of the radiation continuum. These modes contribute primarily to the nonresonant background and improve the completeness of the linear-field expansion. These modes are the seven modes with imaginary part greater than 20. While the two modes with low imaginary part are the modes QNM\(_1\) and QNM\(_2\).

\section{Modes with non-zero Floquet wave vector}

\begin{figure}[H]
    \centering
    \includegraphics[width=0.75\linewidth]{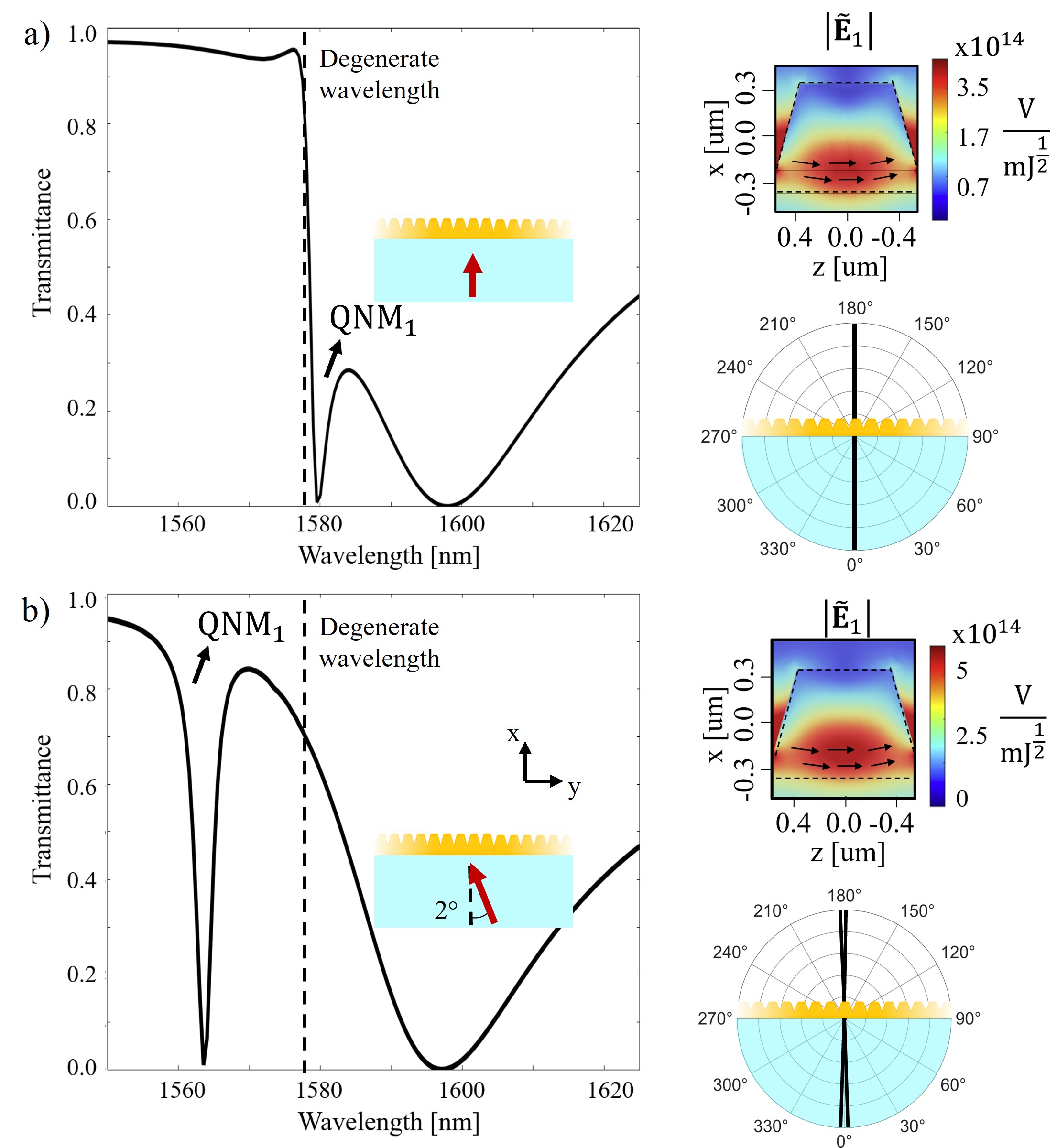}
    \caption{\textit{Representative angle-dependent response of the QNM$_1$ resonance branch.}
    (a) Linear transmittance spectrum under normal incidence, corresponding to
    $\mathbf{k}_{\parallel}=0$, together with the near-field and far-field
    radiation pattern of the associated $\Gamma$-point QNM.
    (b) Linear transmittance spectrum under oblique incidence at
    $\theta=2^\circ$, together with the QNM calculated using the corresponding
    nonzero Floquet wave vector. The near-field profile remains similar to that
    of QNM$_1$, indicating that the two modes belong to the same resonance
    branch, while the resonance wavelength and radiative channels are modified
    at nonzero $\mathbf{k}_{\parallel}$.}
    \label{fig:NonDegenerate}
\end{figure}

In the main text, the QNM analysis is restricted to the $\Gamma$-point resonance QNM$_1$, obtained with the COMSOL eigenfrequency solver by imposing a zero Floquet wave vector, $\mathbf{k}_{\parallel}=0$. This corresponds to the normal-emission channel of the infinite periodic metasurface and provides the reduced mode-resolved description used in the main analysis to identify the physical mechanisms governing photon-pair generation and collection.

In a periodic metasurface, however, the QNM eigenfrequencies and radiative properties generally depend on the transverse wave vector. To illustrate this dependence, we performed angle-dependent linear-transmittance simulations and QNM calculations at nonzero Floquet wave vector. Fig. ~\ref{fig:NonDegenerate}(a) shows the
transmittance spectrum at normal incidence, where the narrow resonance dip is associated with QNM$_1$. Fig. ~\ref{fig:NonDegenerate}(b) shows the corresponding response for oblique incidence at $\theta=2^\circ$. In this case, the resonance is shifted toward shorter wavelengths, consistently with the angle-dependent transmittance
reported in the Supplementary Material of the original experimental work~\autocite{santiago2021photon}.

The corresponding QNMs, calculated at $\mathbf{k}_{\parallel}=0$ and at the nonzero Floquet wave vector associated with $\theta=2^\circ$, are shown on the right-hand side of Fig.~\ref{fig:NonDegenerate}. The off-$\Gamma$ mode exhibits a near-field distribution similar to that of QNM$_1$, indicating that it belongs to the same resonance branch. At the same time, its complex eigenfrequency is shifted and its far-field radiation is redirected
toward oblique channels.

This example shows that restricting the QNM description to $\mathbf{k}_{\parallel}=0$ constitutes a reduced description of the periodic metasurface response. In a more general scenario, particularly when the strongest resonance is not at the degenerate wavelength with  $\mathbf{k}_{\parallel}=0$, QNMs at nonzero $\mathbf{k}_{\parallel}$ may become particularly relevant when their resonance wavelength approaches the degenerate SPDC wavelength, thereby minimizing the detuning term in Eq.~(2) of the main text. A complete treatment of SPDC would therefore require the consistent inclusion of the relevant QNMs at nonzero $\mathbf{k}_{\parallel}$, together with their normalization, nonlinear modal overlaps, and radiative properties.

\section{Finite size metasurface}
\begin{figure}[H]
    \centering
    \includegraphics[width=0.85\linewidth]{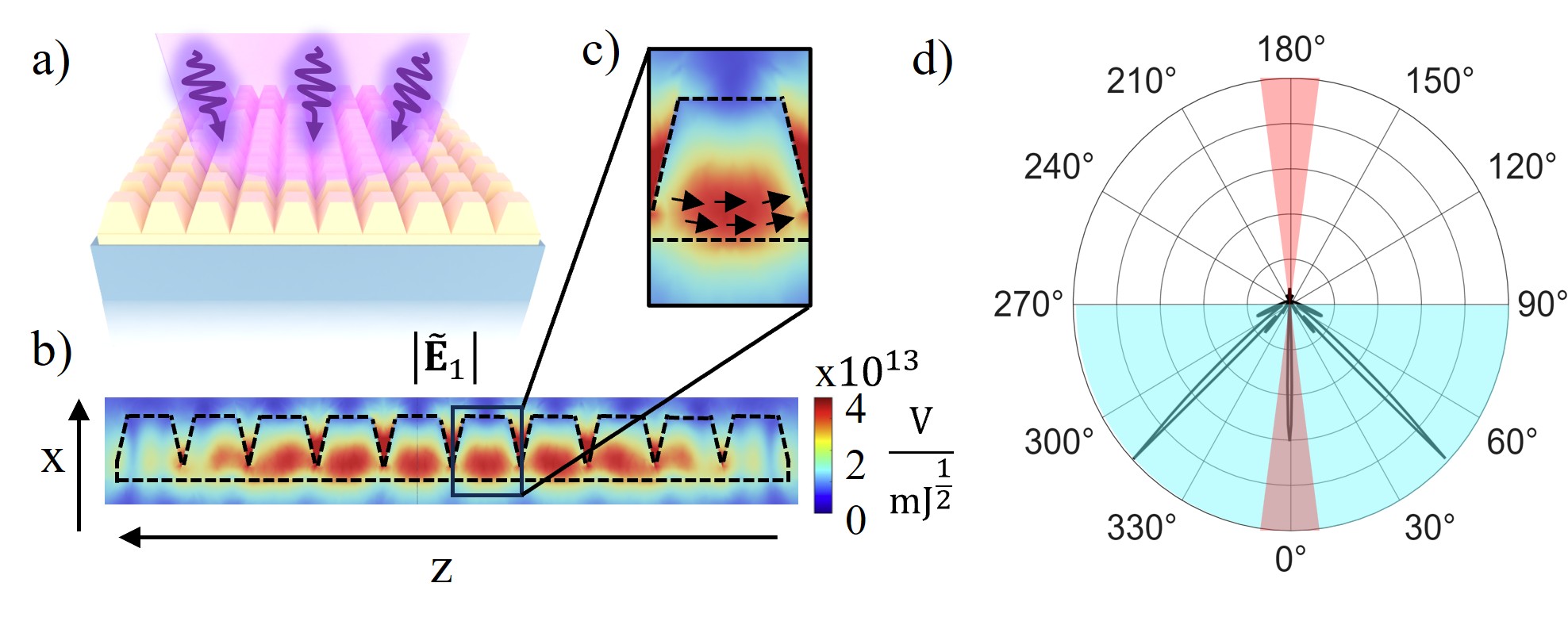}
    \caption{\textit{Change of emission direction with a finite pump width} (a) The metasurface studied constis of only 10 \(\times\) 10 unit-cells, simulating the effect of the experimental pump beam of 6 \(\mu\)m. The electric field of the resonance in the finite case is shown in (b) and in (c) a magnified view of the central unit-cell is shown. (d) The far-field of the corresponding resonance is shown, the blue area indicates the substrate while the red area indicates the angle that is acquired in the original experiment, determined by the numerical aperture of the fiber.}
    \label{fig:gaussian}
\end{figure}

To assess the impact of the infinite-array approximation, we investigated the emission from a finite metasurface in which only a few unit cells are effectively excited, as schematically shown in Fig.~\ref{fig:gaussian} (a). This configuration mimics the experimental situation, where the finite pump waist illuminates only a limited number of resonators.

Fig.~\ref{fig:gaussian} (b) shows the near-field distribution of the corresponding finite-array mode, \(|\tilde{\mathbf E}_1|\). The field retains the main features of the infinite-periodic QNM, with a strong electric field localized inside the LN resonators and predominantly polarized along the optical axis, i.e. the \(z\)-direction. This behavior is particularly clear in the central unit cells, shown in the zoom of Fig.~\ref{fig:gaussian} (c), which are less affected by the termination of the finite array.

The corresponding far-field emission pattern is shown in Fig.~\ref{fig:gaussian} (d), calculated according to \autocite{yang2016near}. In contrast to the infinite-periodic case, where the array factor strongly constrains the emission close to the metasurface normal, the finite array exhibits a trilobate angular spectrum with two lateral lobes in addition to the central one. This broadening arises from the relaxation of in-plane momentum selection and from radiation leakage associated with the finite edges of the structure. As a result, the mode can radiate over a wider range of angles.

However, most of the high-angle radiation lies outside the numerical aperture of the experimental collection system, indicated by the red shaded region in Fig.~\ref{fig:gaussian} (d). Therefore, only the emission close to the normal direction contributes significantly to the detected signal. In this angular range, the finite array shows a pronounced asymmetry between the air and substrate half-spaces. This asymmetry originates from the combined effect of the finite excitation aperture and the refractive-index mismatch between the air and glass sides. Consequently, the emission collected from the air side is strongly suppressed compared to the substrate side, explaining why no measurable signal was observed experimentally in the air-side detection configuration \autocite{santiago2021photon}.

\section{Substrate effects induced by diffraction channels}

\begin{figure}[H]
    \centering
    \includegraphics[width=\linewidth]{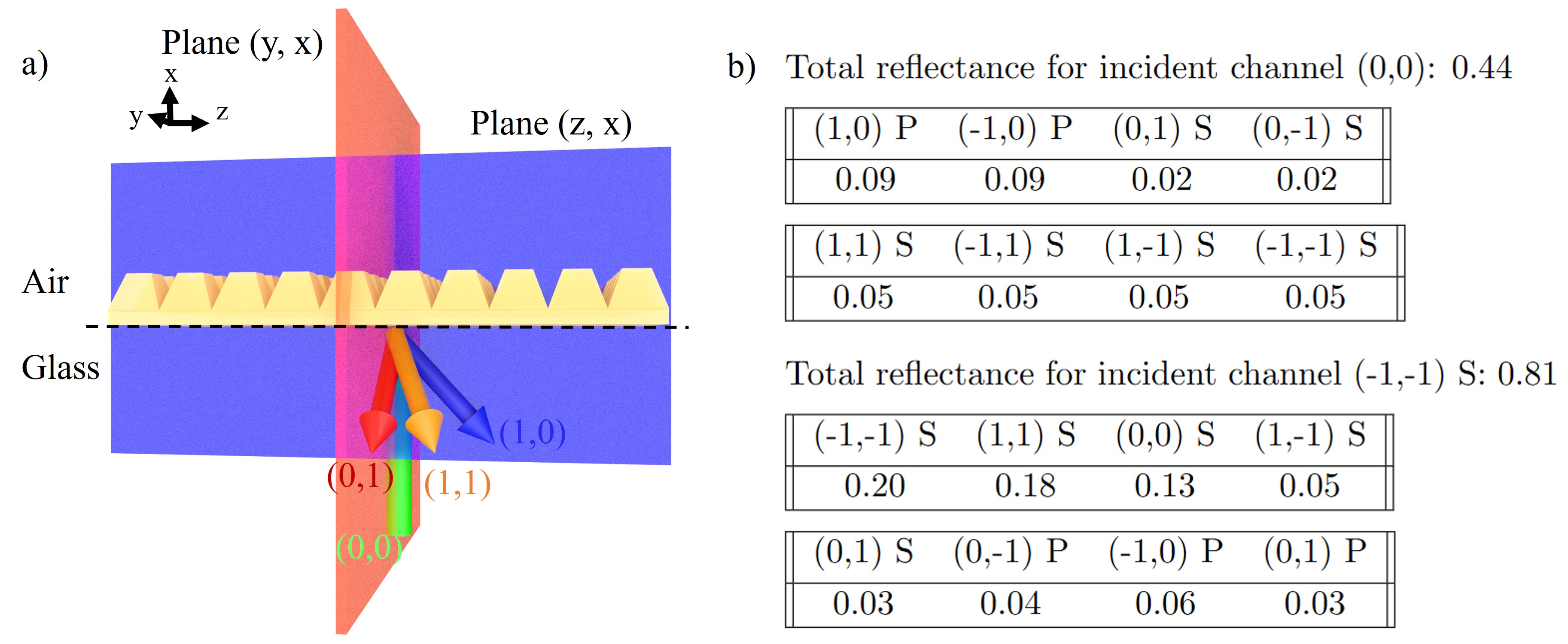}
    \caption{\textit{Pump diffraction orders generated by the metasurface.} (a) Schematic representation of the relevant pump diffraction orders inside the substrate. (b) Summary of the main reflectance coefficients associated with the most relevant diffraction channels.}
    \label{fig:DiffractionOrders}
\end{figure}

The oscillatory behavior of the SPDC efficiency as a function of the substrate thickness can be understood as a finite-substrate interference effect involving the pump diffraction orders generated by the metasurface. Since the metasurface period is \(900~\mathrm{nm}\), larger than the pump wavelength inside the glass substrate, the normally incident pump can be scattered into several propagating diffraction channels.

Fig. ~\ref{fig:DiffractionOrders} (a) shows a schematic representation of the dominant diffraction orders generated inside the glass substrate. The zeroth-order channel \((0,0)\) is shown in green, while three representative diffracted orders are highlighted in red \((0,1)\), blue \((1,0)\), and orange \((1,1)\). For normal incidence, the total reflected power is approximately \(0.44\), while the remaining power is transmitted through the structure. The corresponding highest reflectance coefficients for the different diffraction orders and for polarizations parallel (\(P\)) and perpendicular (\(S\)) to the diffraction plane are reported in the upper table of Fig.~\ref{fig:DiffractionOrders} (b).

Owing to the symmetry of the metasurface under normal incidence, diffraction orders propagating in opposite transverse directions, such as  \((-1,0)\) and \((0,-1)\), exhibit the same reflectance coefficients. The dominant reflected contributions are associated with the \((\pm1,0)\) diffraction orders with \(P\) polarization, which propagate in the corresponding diffraction plane at an polar angle of approximately \(37^\circ\) with respect to the surface normal. Weaker contributions are associated with the \((0,\pm1)\) orders with \(S\) polarization.

In addition, diffraction into the diagonal orders \((\pm1,\pm1)\) occurs at larger propagation angles, approximately \(59^\circ\). Since this angle is larger than the critical angle at the glass--air interface, these orders undergo total internal reflection and are redirected back toward the metasurface. In contrast, for the diffraction orders propagating at $\theta \approx 37^\circ$, the field is mainly transmitted through the glass-air interface for $P$ polarization, while only about 0.2 of the incident power is reflected for $S$ polarization.

To further investigate this mechanism, we calculated the diffraction response of the metasurface for an incident diagonal diffraction channel corresponding to the field of a diagonal order reflected back by the glass--air interface. The resulting highest diffraction efficiencies are reported in the lower table of Fig.~\ref{fig:DiffractionOrders} (b), where the orders are expressed in the global reference system. In this case, most of the incident power is reflected, with a total reflectance of approximately \(0.81\). The dominant reflected channels correspond to the same diagonal order and to the symmetry-related/specular diagonal channel, while a non-negligible contribution is also scattered into the zeroth order, that, however, is mainly transmitted in the glass--air interface.

\begin{figure}
    \centering
    \includegraphics[width=0.75\linewidth]{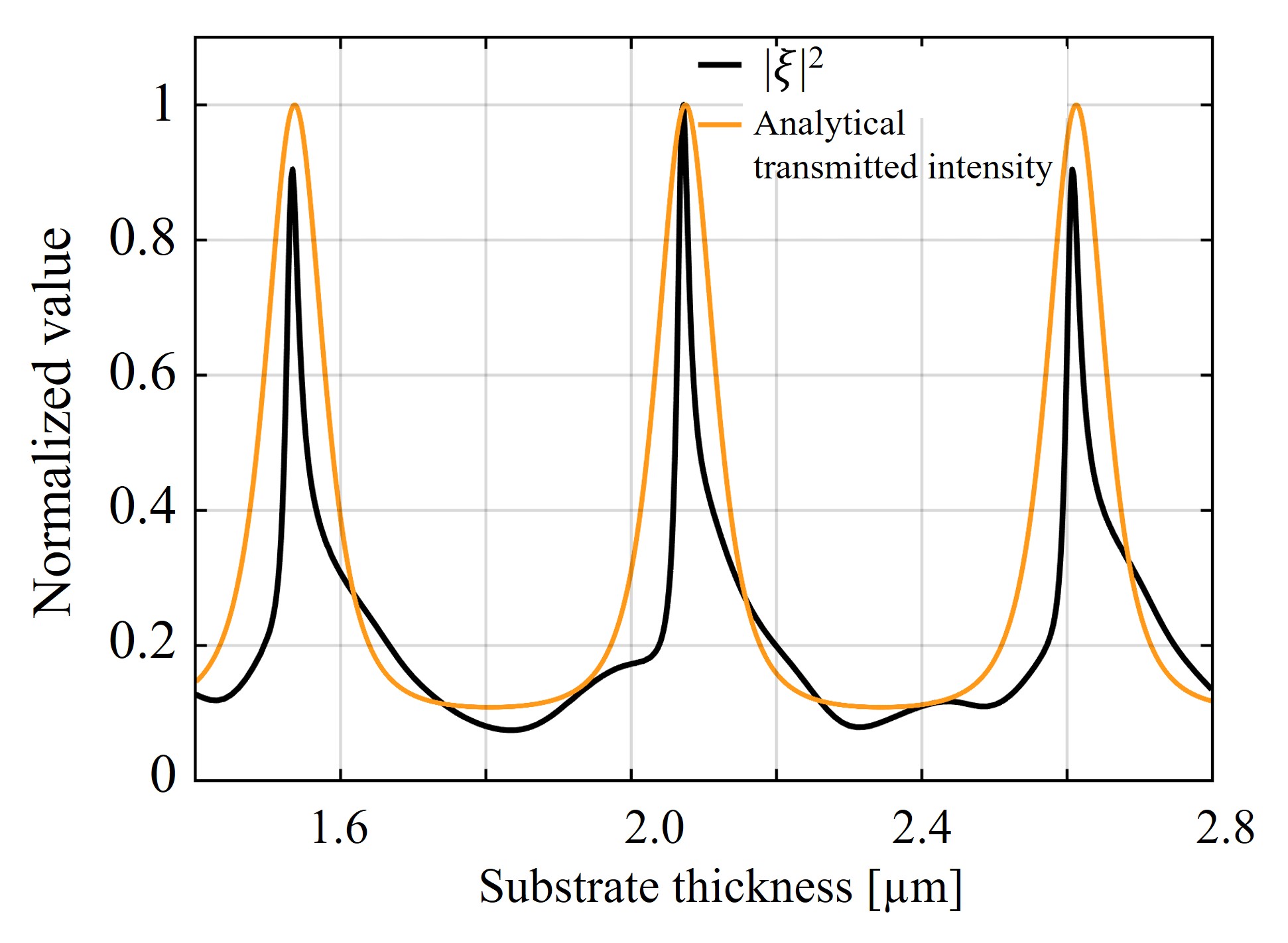}
    \caption{\textit{Comparison between the simulated photon-pair generation efficiency and the analytical transmitted pump intensity.}
    The analytical transmitted intensity is obtained by considering the dominant pump diffraction orders and accounting for their propagation and multiple reflections inside the finite substrate. In particular, the model includes the main contribution from the diagonal diffraction orders \((\pm1,\pm1)\), which are responsible for the observed substrate-thickness-dependent modulation.}
    \label{fig:Analytical}
\end{figure}

These results show that the pump field dynamics inside the finite substrate are dominated by multiple reflections of the diffracted diagonal pump orders. Depending on the substrate thickness, the different reflected contributions acquire different propagation phases and interfere constructively or destructively at the metasurface position. This modifies the pump field coupled into the LN resonators and gives rise to the oscillatory behavior of the nonlinear overlap \(|\xi|^2\), as shown in Fig.~\ref{fig:Analytical}.

Based on these dominant diffraction channels, we constructed a simplified analytical model in which we retrieve the transmitted intensity of the pump as a function of the substrate thickness. The resulting transmitted pump intensity inside the nonlinear material is shown in Fig.~\ref{fig:Analytical} in the orange curve while the black curve is the \(|\xi|^2\) shown in the main text. The model reproduces the main oscillation period observed in the full numerical simulations, confirming that the substrate-thickness dependence originates from the interference of diffracted pump orders.

The oscillation period is governed by the phase accumulated by a diffracted order during a round trip inside the substrate. For a diffraction order propagating at an angle \(\theta\) inside the glass, this phase factor is
\begin{equation}
e^{2 \, i \, k_0 \, n_{\mathrm{glass}} \, d \, \cos\theta},
\end{equation}
where \(k_0=2\pi/\lambda_\mathrm{p}\), \(n_{\mathrm{glass}}\) is the refractive index of the substrate, and \(d\) is the substrate thickness. Constructive interference is therefore obtained when
\begin{equation}
2 \, k_0 \, n_{\mathrm{glass}} \, d \, \cos\theta = 2 \, \pi \, m ,
\end{equation}
with \(m\) an integer. The corresponding thickness period is
\begin{equation}
T =
\frac{\lambda_\mathrm{p}}{2 \, n_{\mathrm{glass}} \, \cos\theta}.
\end{equation}
For the dominant diffraction channel considered here, this gives
\begin{equation}
T \simeq 573~\mathrm{nm},
\end{equation}
in agreement with the periodicity observed in the numerical simulations. The small shift of the peak positions and the residual modulations arise from additional diffraction channels, phase shifts upon reflection, and higher-order multiple-reflection processes that are not fully included in the simplified model.

For a more complete characterization, and to enable future quantitative comparisons, a full transfer-matrix-method approach could be implemented, following related studies of SHG in multilayer configurations \autocite{pakhomov2021modeling, wang2024theory}.

\section{Proof-of-principle multilayer substrate design}

\begin{figure}[H]
    \centering
    \includegraphics[width=0.95\linewidth]{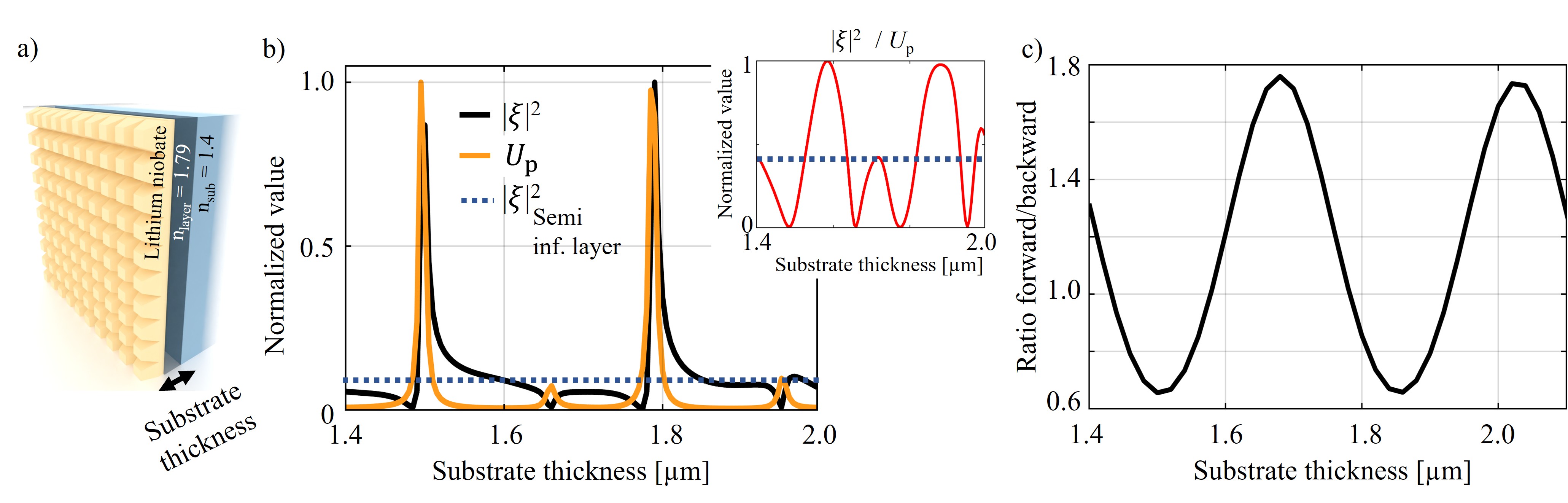}
    \caption{\textit{Preliminary multilayer substrate design.}(a) Schematic of the multilayer geometry, consisting of an LN metasurface on a finite dielectric layer with refractive index \(n\simeq 1.79\), followed by a semi-infinite glass-like substrate with \(n\simeq 1.4\). The finite layer thickness is varied.  (b) Normalized nonlinear-overlap contribution \(|\xi|^2\) and integrated pump-field intensity \(U_\mathrm{p}\) as a function of the finite-layer thickness. The blue dashed line indicates the corresponding semi-infinite-substrate reference value. The inset shows the normalized ratio \(|\xi|^2/U_\mathrm{p}\). (c) Ratio between the forward and backward direction of the photon pairs as a function of the substrate thickness.}
    \label{fig:multilayer}
\end{figure}
In the main text, we considered a finite air-backed glass substrate as an idealized geometry to isolate the physical mechanism responsible for substrate-induced enhancement, namely the recoupling and interference of pump diffraction orders. Here, we examine a preliminary multilayer configuration aimed at demonstrating that the same mechanism can, in principle, be implemented in a more realistic substrate geometry. This example is not intended to represent a fully optimized device, but rather a proof-of-principle demonstration of multilayer substrate engineering for enhancing the nonlinear overlap.

For realistic implementations, an important constraint must be considered. The diffraction orders responsible for the recoupling mechanism propagate obliquely inside the layer below the metasurface. Therefore, the reflecting or recoupling interface must be sufficiently close to the metasurface so that the reflected diffraction orders still spatially overlap with the illuminated region. In macroscopic substrates, the lateral displacement of these oblique orders can become much larger than the pump-beam waist, strongly reducing their ability to recouple to the metasurface. This motivates the use of a finite optical layer with a thickness of the order of a few micrometers, followed by an additional substrate or multilayer structure.

The multilayer geometry considered here is shown in Fig.~\ref{fig:multilayer} (a). The LN metasurface is deposited on a finite dielectric layer with refractive index \(n_{\text{layer}}\simeq 1.79\), representative of a sapphire-like material, followed by a semi-infinite glass-like substrate with refractive index \(n_{\text{sub}}\simeq 1.4\). This configuration preserves a finite optical path below the metasurface while avoiding the idealized air-backed substrate considered in the main text.

We emphasize that, in this proof-of-principle example, the resonator geometry was redesigned with respect to the original telecom-wavelength metasurface. The purpose of this redesign is to place the relevant pump diffraction orders in a regime where they can be internally reflected at the interface between the finite dielectric layer and the lower-index substrate. For a diffraction order \((m,n)\), the propagation angle inside the finite layer is determined by
\begin{equation}
\sin\theta_\mathrm{mn}
=
\frac{\lambda_{\mathrm{p}}}{n_{\mathrm{layer}}\,p}
\sqrt{m^2+n^2},
\end{equation}
where \(\lambda_{\mathrm{p}}\) is the pump wavelength and \(p\) is the metasurface period. Internal reflection at the interface with the lower-index substrate requires
\begin{equation}
\sin\theta_\mathrm{mn}>
\frac{n_{\mathrm{sub}}}{n_{\mathrm{layer}}}.
\end{equation}
Therefore, efficient multilayer-enhanced recoupling is not automatic, but requires simultaneous engineering of the lattice period, pump wavelength, and refractive-index contrast.

To satisfy this condition, the resonator geometry was redesigned to support the dominant QNM around the degenerate wavelength \(\lambda_\mathrm{s}=\lambda_\mathrm{i}\simeq 1250~\mathrm{nm}\), corresponding to a pump wavelength \(\lambda_\mathrm{p}\simeq 625~\mathrm{nm}\). The optimized truncated-pyramid geometry has period \(615~\mathrm{nm}\), height \(864~\mathrm{nm}\), a bottom-to-top base ratio of \(0.64\), and a residual layer thickness of \(420~\mathrm{nm}\). This geometry is used as a representative example rather than as a fully optimized structure. Similar multilayer-enhancement strategies could in principle be implemented at other wavelengths by redesigning the resonator geometry, choosing a different nonlinear material platform, or engineering a suitable substrate stack.

The results are shown in Fig.~\ref{fig:multilayer} (b). As in the finite-substrate analysis discussed in the main text, the nonlinear-overlap contribution \(|\xi|^2\) exhibits pronounced maxima as a function of the finite-layer thickness. These maxima closely follow the peaks of the integrated pump-field intensity inside the nonlinear material, \(U_{\mathrm{p}}\), confirming that the enhancement is mainly associated with constructive recoupling of the pump field into the LN resonators. Compared to the reference semi-infinite substrate with refractive index \(n=1.79\), chosen by having the same resonance frequency, indicated by the blue dashed line, the multilayer configuration provides an enhancement of more than one order of magnitude in \(|\xi|^2\).

Importantly, the enhancement is not solely due to a larger pump-field intensity inside the nonlinear material. As shown in the inset of Fig.~\ref{fig:multilayer} (b), the normalized ratio \(|\xi|^2/U_{\mathrm{p}}\) also exhibits clear thickness-dependent maxima. A maximum improvement of approximately a factor of two is obtained compared to the corresponding semi-infinite reference case. This indicates that the multilayer substrate can improve the coherent nonlinear overlap, and not only the amount of pump field coupled into the LN resonators.

Furthermore, Fig.~\ref{fig:multilayer} (c) shows that the multilayer geometry also induces oscillations in the forward/backward emission ratio. These oscillations arise from etalon-like interference effects inside the finite dielectric layer, which modify the relative photon-pair emission into the two directions.

\printbibliography